\documentstyle[12pt]{article}
\newcommand{\text }{{\rm d}}
\newcommand\bm[1]{\mbox{\boldmath$#1$}}
\begin{document}
\title{General non-rotating perfect-fluid solution with an abelian
spacelike $C_3$ including only one isometry.
}
\author{Andreas Koutras\footnote{e-mail:A.Koutras@qmw.ac.uk} \ and Marc
Mars\footnote{e-mail:M.Mars@qmw.ac.uk}
\thanks{Also at Laboratori de F\'{\i}sica Matem\`atica, Societat
Catalana de F\'{\i}sica, IEC, Barcelona.}
\\
School of Mathematical Sciences,
Queen Mary and Westfield College, \\
Mile End Road, London E1 4NS, United Kingdom }
\maketitle
\begin{abstract}
The general solution for non-rotating perfect-fluid spacetimes
admitting one Killing vector and two conformal (non-isometric)
Killing vectors spanning an abelian three-dimensional conformal algebra ($C_3$)
acting on spacelike hypersurfaces is presented. It is of Petrov
type D; some properties of the
family such as matter contents are given. This
family turns out to be an extension of a solution recently given in
\cite{SeS}
using
completely different methods.
The family contains Friedman-Lema\^{\i}tre-Robertson-Walker 
particular cases
and could be useful as a test for the
different FLRW perturbation schemes.
There are two very interesting limiting cases,
one with a non-abelian $G_2$ and another with an abelian $G_2$
acting non-orthogonally
transitively on spacelike surfaces and with the fluid velocity non-orthogonal
to the group orbits. 
No examples are known to the authors
in these classes.
\end{abstract}

Very few exact perfect fluid solutions of Einstein field equations with
a low degree of symmetry are known even though 
they may prove important for the study of inhomogeneities in the Universe or
in parts of the Universe.
The high complexity of the inhomogeneities of the real Universe
makes the modeling of such structures using exact solutions intractable.
Thus, perturbation schemes (see e.g. \cite{PS1},\cite{PS} and references
therein) are usually
used in order to understand
the evolution of such inhomogeneities (in a 
Friedmann-Lema\^{\i}tre-Robertson-Walker background). However, these
perturbation schemes smooth out any unexpected
behaviours due to the high non-linearity of the theory of general
relativity. These possible new behaviours can only be fully understood by 
finding and analyzing exact solutions of Einstein field equations. Moreover,
they can be significant in testing the ranges in which the
linear approximation of the perturbation theory is valid. In order to
accomplish these objectives 
it is necessary and convenient to find an increasing number of exact
solution with the
less possible
symmetries. In the recent past,
the use of conformal Killing vectors has proved useful in finding new
families of exact solutions with a two-dimensional isometry group (the
so-called $G_2$ cosmologies). For instance, the general
perfect-fluid solution with a two-dimensional isometry group
acting orthogonally transitively  \cite{Wai} on spacelike orbits and admitting
one conformal Killing vector is known \cite{Ali}, \cite{MW}. The next natural
step is trying to determine the perfect-fluid solutions when
the spacetime admits {\it one} isometry and two conformal Killing vectors 
(all of them being spacelike).
The simplest case is when this three-dimensional conformal algebra is
abelian. Since we are interested in cosmological models we will also impose
that the fluid velocity is non-rotating (although rotating cosmological
models are also of great interest, their study is significantly more difficult
than the non-rotating one) and that $\rho+p$ is positive at least in an
open region ($\rho$ and $p$ being the energy-density and pressure
of the fluid respectively). In this letter we present the general solution
of Einstein
field equations under these assumptions\footnote{We  exclude the
conformally flat families since all of them are known \cite{KRM}}.

The spacetimes admitting an abelian $C_3$ algebra of
conformal Killing vectors, with one Killing vector 
and two conformal Killing vectors (CKVs) acting on 
spacelike hypersurfaces, admit coordinates (using Defrise-Carter 
theorem \cite{DC}) in which the line-element reads 
\begin{equation}
\text s^2=\Omega^2(y,z,t)\left[ 
-\frac{\text  t^2}{M(t)} +M(t)\text  x^2+N(t)
\text  y^2+S(t)\text z^2\right] \label{eq:1}
\end{equation}
where $M$, $N$ and  $S$ are arbitrary functions of $t$ and the conformal 
factor $\Omega$ is an arbitrary function of $t$, $y$ and $z$.
The metric (\ref{eq:1}) admits one Killing vector $\partial_x$ and two CKVs
$\partial_y$ and $\partial_z$. The main result of this letter is 

\vspace{5mm}

{\it The 
general non-rotating non-conformally flat perfect-fluid solution of Einstein's
field equations
(having $\rho+p >0$ somewhere) with one Killing and two conformal
Killing vectors spanning an abelian Lie algebra acting 
on spacelike
hypersurfaces is 
\begin{equation}
\text s^2 = 
\frac{N^{\frac{1-\alpha}{\alpha}}(y)}{Q^{\frac{1+\alpha}{\alpha}}(z)}
\left [ - \frac{\text t^2}{A(t)} + A(t) \text x^2 + t^{1+\alpha} 
\text y^2 + t^{1-\alpha} \text z^2
\right ] \label{eq:2}
\end{equation}
where $\alpha$ is a non-vanishing constant, the function $A(t)$ reads
\begin{equation}
A(t) = r_0 + r_1 t^{1-\alpha} + r_2 t^{1+\alpha}
\end{equation}
($r_0$, $r_1$ and $r_2$ are arbitrary constants), and the
 functions $N(y), Q(z)$
satisfy the following trivial differential equations
\begin{eqnarray*}
\left ( \frac{\text N}{\text y} \right )^2 = \alpha^2 
\left ( r_1 N^2 - v_1 \right ), \hspace{1cm} 
\left ( \frac{\text Q}{\text z} \right )^2 = \alpha^2 
\left ( r_2 Q^2 - v_2 \right ), \hspace{1cm} v_1, v_2 \,\,\mbox{const.}
\end{eqnarray*}}

\vspace{5mm}

Thus, the metric is completely explicit and very simple in form. 
The velocity one-form of the fluid is
\begin{eqnarray}
\bm{u} = - \frac{N^{\frac{1-\alpha}{2\alpha}}}{Q^{
\frac{1+\alpha}{2\alpha}}} \frac{1}{\sqrt{R}}
 \left ( \bm{dt} +  \frac{t N_{,y}}{\alpha N } \bm{dy}
-  \frac{t Q_{,z}}{\alpha Q} \bm{dz} \right ), \label{vuf}
\end{eqnarray}
where $R$ stands for
\begin{eqnarray}
R \equiv r_0 + v_1\frac{t^{1-\alpha}}{N^2} + v_2 \frac{t^{1+\alpha}}{Q^2}.
\label{erre}
\end{eqnarray}
This expression must be strictly positive in order to have a perfect-fluid
spacetime. When $R <0$ the matter contents is a tachyon fluid and $R=0$
represents a null fluid.
For certain values of the
parameters the spacetime has a region where the
matter contents is a perfect fluid, there exists a transition hypersurface
where the fluid becomes null (with pressure, in general) and there is
also a non-empty open region where the perfect fluid is tachyonic. This
kind of behaviour is very common when solving perfect-fluid  Einstein field
equations in non-comoving coordinates (see \cite{MW} for
other explicit examples of this fact). 
>From (\ref{vuf}) we learn that the
fluid is highly tilted with respect to the orbits of the 
conformal group. It is 
orthogonal to the Killing vector $\partial_x$ but it is not
orthogonal to either of the two conformal Killings (and consequently it
is not orthogonal to the three-dimensional conformal orbits). It is 
convenient to define a new function $\tau$ by
\begin{eqnarray*}
\tau \equiv t \frac{N^{\frac{1}{\alpha}}}{Q^{\frac{1}{\alpha}}},
\end{eqnarray*} 
so that the 
fluid velocity one-form can be rewritten in a compact form as 
\begin{eqnarray*}
\bm{u} = - \frac{\bm{d\tau}}{\sqrt{r_0  N^{\frac{1+\alpha}{\alpha}}
Q^{\frac{\alpha-1}{\alpha}} + v_1 \tau^{1-\alpha}+ v_2 \tau^{1+\alpha}}}.
\end{eqnarray*}
This expression shows that the fluid velocity is hypersurface
orthogonal and therefore non-rotating, which is one of our main assumptions.
The hypersurfaces orthogonal to the fluid are given by
$\tau = \mbox{const}$ and therefore $\tau$ is a cosmic time for the
spacetime. The energy density and pressure are (using this new time $\tau$) 
\begin{eqnarray}
\rho = \frac{3}{4} \left (1+\alpha \right)^2 \frac{v_2}{
\tau^{1-\alpha}}+
\frac{3}{4} \left (1-\alpha \right )^2 \frac{v_1}{
\tau^{1+\alpha}} + \frac{r_0}{4}
\left (1-\alpha^2 \right ) \frac{N^{\frac{1+\alpha}{\alpha}}}{\tau^2 
Q^{\frac{1-
\alpha}{\alpha}}} \nonumber  \\
p = - \frac{ \left (1 + \alpha \right)\left (1 + 5\alpha \right ) v_2}{4
\tau^{1-\alpha}}+
\frac{ \left (1-\alpha \right ) \left (5 \alpha -1 \right) v_1}{4
\tau^{1+\alpha}} + \frac{r_0}{4}
\left (1-\alpha^2 \right ) \frac{N^{\frac{1+\alpha}{\alpha}}}{\tau^2
Q^{\frac{1-
\alpha}{\alpha}}}, \label{rhopre}
\end{eqnarray}
which imply
\begin{eqnarray*}
\rho + p = \frac{1}{2}\left (1- \alpha^2 \right) \frac{R N^{
\frac{1+\alpha}{\alpha}}}{\tau^2 Q^{\frac{1-\alpha}{\alpha}}}
\end{eqnarray*}
and therefore the energy condition $\rho+p >0$ is
fulfilled everywhere provided the constant $\alpha$ is
restricted to $\alpha^2 <1$.
The family of solutions is invariant under the simultaneous changes $\alpha
\leftrightarrow -\alpha$, $y \leftrightarrow z$, $N \leftrightarrow Q$.
Thus, we can assume without loss of generality that $\alpha$ is
strictly positive and then the energy condition imposes
\begin{eqnarray*}
0 < \alpha < 1.
\end{eqnarray*}
Expression (\ref{rhopre}) shows that the
spacetime is singular at $\tau=0$ where a big bang singularity occurs. Assuming
$r_0>0$ there always exist a non-empty open region near 
the big bang singularity
where both the density and pressure are positive.
For $v_1$ and $v_2$
non-negative we have $R>0$ everywhere (so that the matter contents is
perfect fluid in the whole spacetime) and the energy-density is positive
everywhere.

The Petrov type of the spacetime is D and 
in a null tetrad adapted to the two repeated null principal directions
\begin{eqnarray*}
\bm{l} =  \frac{1}{\sqrt{2}}\frac{N^{\frac{1-\alpha}{2\alpha}}}{Q^{
\frac{1+\alpha}{2\alpha}}} \left ( \frac{\bm{dt}}{\sqrt{A}}+
 \sqrt{A} \bm{dx} \right ), \hspace{1cm}
\bm{k} = \frac{1}{\sqrt{2}}\frac{N^{\frac{1-\alpha}{2\alpha}}}{Q^{
\frac{1+\alpha}{2\alpha}}} \left ( \frac{\bm{dt}}{\sqrt{A}}-
 \sqrt{A} \bm{dx} \right ),
\end{eqnarray*}         
the only non-vanishing Weyl
spinor component reads
\begin{eqnarray}
\Psi_2 = \frac{r_0 \left (\alpha^2 -1 \right) Q^{\frac{1+\alpha}{\alpha}}}{
12 t^2 N^{\frac{1-\alpha}{\alpha}}}. \label{Psi}
\end{eqnarray}
These expressions show that the fluid velocity does not lie in the two-plane
generated at each point by the repeated null directions. Both
null directions are geodesic and non-rotating and they are
shearing and expanding. 
Furthermore, the acceleration of the fluid $\bm{\dot{u}}$
does not lie in the plane generated by $\bm{l}$ and $\bm{k}$ (see 
{\cite{KRM}).

>From (\ref{Psi}) we have that
the conformally flat subcases of the solution are obtained when either
$\alpha=1$ or $r_0=0$. The metric with $\alpha=1$ is de Sitter ($v_2>0$),
anti-de Sitter ($v_2 <0$) or Minkowski ($v_2=0$). When $r_0=0$ (arbitrary
$\alpha$) the condition $R>0$ (\ref{erre})
implies that at least one of the $v_1$, $v_2$ must be positive. The fluid
satisfies a barotropic equation of state (\ref{rhopre}) and therefore
the spacetime must be a Friedmann-Lema\^{\i}tre-Robertson-Walker (FLRW)
cosmology (see e.g. \cite{WL}).

It is remarkable that the general family we present in this letter is
an extension of a solution recently found in \cite{SeS} using completely
different methods. The two families coincide when either $v_1$ or $v_2$
are positive and, therefore, the solutions with $v_1 \leq 0$ and $v_2 \leq 0$
presented here are new. In \cite{SeS}, the authors use the
Kerr-Schild
transformation to find perfect-fluid solutions starting from a
FLRW  seed metric. This seed metric is exactly the 
subcase $r_0=0$, $v_1>0$ in (\ref{eq:2}) (written in different
coordinates). After performing the Kerr-Schild transformation, they find
a family of solutions which is equivalent\footnote{The two coordinate systems
are different and therefore it is not obvious that
the two families coincide.} to the subfamily
$v_1 >0$ (or $v_2>0$) in (\ref{eq:2}).
The solutions with $v_1\leq 0$, $v_2 \leq 0$ were not found using
the Kerr-Schild method in \cite{SeS} because their conformally 
flat limit is not
FLRW. Instead, one gets a tachyon fluid (admitting a $G_6$) when $v_1<0$
and $v_2 \leq 0$ (or equivalently $v_1 \leq 0$ and $v_2<0$) and
a radiation solution admitting $G_7$ when $v_1=v_2=0$.
It can be seen, however,
that the whole family presented here can be generated using the
Kerr-Schild ansatz starting from the seed metric (\ref{eq:2}) with $r_0=0$.
This fact is most
remarkable since the Kerr-Schild transformation and the existence of
conformal Killing vectors have no apparent relationship. Thus, the
family of perfect-fluid solutions (\ref{eq:2}) has two
completely different and apparently unrelated characterizations as
the most general solution with an abelian spacelike $C_3$ including one
isometry  and the most general solution which can be found from 
the metric with $r_0=0$ using the Kerr-Schild transformation. The possible
reason for such an unexpected connection should be a matter for further
investigation. 

The analysis of the Killing equations for the metric (\ref{eq:2})
shows that
there is only one Killing vector $\vec{k_1}= \partial_x$ except
for the three following subcases (apart from $\alpha^2=1$ and $r_0=0$ which
have been discussed above).
\begin{itemize}
\item[A)]
When $r_1=r_2=0$, $v_1<0$ and $v_2<0$ (therefore, this is a new solution
not included in \cite{SeS}) the metric admits a 
non-abelian $G_2$ and the line-element can be written in the form
\begin{eqnarray*}
\text s^2 = D^2 \frac{y^{\frac{1-\alpha}{\alpha}}}{z^{\frac{1+\alpha}{\alpha}}}
\left [ - \text t^2 + \text x^2 + t^{1+\alpha}\text y^2 +
t^{1-\alpha}\text z^2 \right ].
\end{eqnarray*}
(where $D$ cannot be reabsorbed into the coordinates).
This metric has only two Killing vectors
\begin{eqnarray*}
\vec{k_1} = \frac{\partial}{\partial x}, \hspace{1cm}
\vec{k_2} = t \frac{\partial}{\partial t} + x \frac{\partial}{\partial x}+
\frac{1-\alpha    }{2}y \frac{\partial}{\partial y} +
\frac{ 1+\alpha  }{2}z  \frac{\partial}{\partial z} 
\end{eqnarray*}
with commutator
$ \left [ \vec{k_1}, \vec{k_2} \right ] = \vec{k_1}$. 
$\vec{k_1}$ is spacelike everywhere, but
$\vec{k_2}$ changes its spacelike, null and timelike character through the
spacetime.
The fluid velocity also changes its character. It is timelike
(and thus a perfect fluid) in a region near the big-bang and it becomes
spacelike for big enough values of $\tau$. It can be proved that in the region
where the matter contents is perfect fluid (the physical region) the
Killing vector $\vec{k_2}$ is spacelike everywhere. After the
fluid has become tachyonic, $\vec{k_2}$ becomes timelike and thus the
spacetime is stationary. As far as we know, this solution is the first 
explicit exact solution for a perfect fluid with a
non-abelian two-dimensional maximal isometry group (see \cite{VB} for
a discussion of some properties of non-abelian $G_2$ perfect-fluid
spacetimes).
\item [B)]
When $r_1=v_1=0$ (i.e $N_{,y}=0$)  and $Q_{,z}\neq0$ the metric has an 
abelian $G_2$ acting on spacelike orbits. The two Killings are $\partial_x$
and $\partial_y$ (there is a similar case when $Q_{,z}=0$ and $N_{,y} \neq 0$,
then the Killings are obviously $\partial_x$ and $\partial_z$).
\item[C)]
When $v_2=v_1=0$ (then necessarily $r_1$ and $r_2$ must be non-negative
and we can write them as $r_1= c_1^2$, $r_2 = c_2^2$). The metric 
takes the form
\begin{eqnarray}
\text s^2 &= &
e^{\left (1-\alpha \right) c_1 y - \left (1+\alpha \right) c_2 z}
\left [ -\frac{\text t^2}{r_0+c_2^2 t^{1+\alpha}+ c_1^2 t^{1-\alpha}}
+\left (r_0+c_2^2 t^{1+\alpha}+ 
c_1^2 t^{1-\alpha} \right ) \text x^2 + \right .\nonumber  \\
& & \left .\vspace{9mm}  +
t^{1+\alpha} \text y^2 + t^{1-\alpha} \text z^2 \right ], \label{metrc1c2}
\end{eqnarray}
and the perfect-fluid satisfies a barotropic
equation of state for a stiff fluid
\begin{eqnarray*}
p = \rho = \frac{r_0 \left (1- \alpha^2 \right)}{4 t^2}
e^{\left (\alpha -1 \right) c_1 y + \left (1+\alpha \right) c_2 z} > 0.
\end{eqnarray*}
Since this metric has $v_1=v_2=0$ it is not contained in the family
previously found in \cite{SeS}. 
This metric has two Killing vectors, $\vec{k_1}$ and $\vec{k_2}$,
and one homothety $\vec{h}$ (assuming $c_1$ or $c_2$ non-zero, 
otherwise the metric is a Bianchi I cosmology) given by
\begin{eqnarray*}
\vec{k_1} = \frac{\partial}{\partial x}, \quad
\vec{k_2}= \left (1\!+\!\alpha \right) c_2 \frac{\partial}{\partial y} \!+\!
 \left (1\!-\!\alpha \right) c_1 \frac{\partial}{\partial z}, \quad
\vec{h} =  
\left (1\!+\! \alpha \right) c_1 \frac{\partial}{\partial y}
\!- \!\left (1\!-\! \alpha \right)c_2 \frac{\partial}{\partial z}
\end{eqnarray*}
which are commuting and spacelike everywhere. The one-form associated
with $\vec{k_1}$ is clearly integrable while the one-form associated
with $\vec{k_2}$ is  not only non-integrable (assuming  $c_1$ and 
$c_2$ non-zero, otherwise  the metric is a so-called diagonal
cosmology \cite{Wai} ) but also satisfies
\begin{eqnarray*}
\bm{k_1} \wedge \bm{k_2} \wedge \bm{d k_2} \neq 0,
\end{eqnarray*}
which means that the orbits of the two-dimensional group are not surface
orthogonal and thus the isometry group is not orthogonally transitive. In
addition the fluid velocity is not orthogonal to the isometry group orbits.
This
solution belongs to $A(\mbox{ii})$ in Wainwright's classification of $G_2$
cosmologies
\cite{Wai}.
Very few exact solutions in this class are known and to the best of
our knowledge all of them have the fluid velocity orthogonal to the
isometry group orbits (\cite{Wil}, \cite{VB2}). This is a general
property for B(i) and B(ii) classes, but it is an added assumption for
A(i) and A(ii) classes. In our case
the fluid velocity
is {\it not} orthogonal to the isometry group orbits and as far as we know
this is
the first example with this property. 
\end{itemize}

As a final comment, let us emphasize that the family of solutions (\ref{eq:2})
is given explicitly
in terms of elementary
functions and is of very simple form, despite the low isometry group.
Thus, it may be suitable for
testing the range of validity of
the different FLRW perturbation schemes. In addition, the $r_0$ parameter
controls the deviation from FLRW in a very neat way and hence
any possible disagreement between the exact model and the predictions
of the perturbation schemes can easily be detected and interpreted.      

\section*{Acknowledgements}
This work was partly supported by NATO Science Fellowship.
M.M. wishes to thank Ministerio de Educaci\'on y Ciencia for financial
support under grant EX95 40985713.


\begin{thebibliography}{999}
\bibitem{PS1} P.Coles, F.Lucchin, (1996) {\it Cosmology: the origin
and evolution of cosmic structure}, John Wiley \& Sons, New York.
\bibitem{PS} G.F.R.Ellis in {\it Inhomogeneous cosmological models} (1995),
World
Scientific, A.Molina \& J.M.M.Senovilla Eds.
\bibitem{Wai} J.Wainwright, (1981) {\it J.Phys. A: Math. Gen} {\bf 14} 1131.
\bibitem{Ali} J.Carot, A.A.Coley, A.M.Sintes, (1996) {\it Gen. Rel. Grav.}
{\bf 28} 311.
\bibitem{MW}  M.Mars, T. Wolf, (1996) submitted to {\it Class. Quantum
Grav.}
\bibitem{WL} G.F.R. Ellis in {\it General Relativity and Cosmology},
Proceedings
of the international school of physics Enrico Fermi (1971), Academic Press,
New York. 
\bibitem{KRM} D.Kramer, H.Stephani, M.MacCallum, E.Herlt, (1980) {\it Exact
solutions of Einstein's field equations}, Cambridge Univ. Press, Cambridge.
\bibitem{DC} L.Defrise-Carter, (1975) {\it Commun. Math. Phys.} {\bf 40} 273.
\bibitem{SeS} J.M.M.Senovilla, C.F.Sopuerta, (1994), {\it
Class. Quantum Grav.} {\bf 11} 2073.
\bibitem{VB} N. Van den Bergh, (1988) {\it Class. Quantum Grav.} {\bf 5}
861.
\bibitem{Wil} P.Wils, (1991) {\it Class. Quantum Grav.} {\bf 8} 361. 
\bibitem{VB2} N.Van den Bergh, (1988) {\it Class. Quantum Grav.} {\bf 5}
167.
\end{thebibliography}
\end{document}